\documentclass[onecolumn]{pasj00}
\begin{document}
\SetRunningHead{S. Urakawa et al.}
{An Extrasolar Planet Transit Search with Subaru Suprime-Cam}
%\Received{2005/08/20}%{yyyy/mm/dd}
%\Accepted{2001/01/01}%{yyyy/mm/dd}

\title{Extrasolar Transiting Planet Search with Subaru Suprime-Cam}

%%% begin:list of authors

\author{Seitaro \textsc{Urakawa}\altaffilmark{1}, Toru
\textsc{Yamada}\altaffilmark{2}, Yasushi
\textsc{Suto}\altaffilmark{3}, Edwin
L. \textsc{Turner}\altaffilmark{4}, \\Yoichi
\textsc{Itoh}\altaffilmark{1}, Tadashi
\textsc{Mukai}\altaffilmark{1}, Motohide
\textsc{Tamura}\altaffilmark{2} and Yiping
\textsc{Wang}\altaffilmark{5}

}

\altaffiltext{1}{ Graduate School of Science and Technology, Kobe University,\\
 1-1 Rokkodai-cho, Nada-ku, Kobe, Hyogo 657-8501}
\email{urakawa@kobe-u.ac.jp,yitoh@kobe-u.ac.jp, mukai@kobe-u.ac.jp}
\altaffiltext{2}{National Astronomical Observatory of Japan, 2-21-1 Osawa, Mitaka, Tokyo 181-8588}
\email{yamada@optik.mtk.nao.ac.jp,hide@optik.mtk.nao.ac.jp}
\altaffiltext{3}{Department of Physics, School of Science, The University of Tokyo, Tokyo 113-0033}
\email{suto@utap.phys.s.u-tokyo.ac.jp}
\altaffiltext{4}{Princeton University Observatory, Peyton Hall, Princeton, NJ 08544,USA}
\email{elt@astro.princeton.edu}
\altaffiltext{5}{Purple Mountain Observatory, Chinese Academy of Sciences, China}
\email{ypwang@pmo.ac.cn}
%%% end:list of authors

%%% Please use the following style in case that sorting by 
%%% affilation is impossible. 
 %Do NOT move this preamble from here!

\maketitle

\begin{abstract}
We report the results of a prototype photometric search for transiting extrasolar
planets using Subaru Suprime-Cam.  Out of about 100,000 stars
monitored around the Galactic plane ($l=90^{\circ}$,$b=0^{\circ}$), we
find that 7,700 (27,000) stars show the photometric precision below 1\%
(3\%) for 60 second exposures which is required to detect extrasolar
planets by the transit method.  Thus Suprime-Cam has the photometric
stability and accuracy required for a transiting planet survey.  During
this observing run, we detected three transiting planetary candidates
(i'-band magnitude around 18.5) which exhibit a single full transit-like 
light curve with a fractional depth of $<5\%$.  While future photometric and/or
spectroscopic follow-ups remain to be done, the estimated parameters for
the three systems are consistent with planetary companions around
main-sequence stars.  We also found two eclipsing binary candidates and
eleven variable stars exhibiting W UMa-like light curves.
\end{abstract}

\section{Introduction}

Since the first discovery in 1995 \citep{key-1}, more than 170
extrasolar planets around main sequence stars 
have been reported so far.  While a majority of them
were discovered by the radial velocity method, the transit method, i.e.,
detecting a small decrement in stellar light due to occultation by
planets, provides a unique complementary means to estimate the precise
radius and orbital inclination of those planets which exhibit transit.
Combined with the radial velocity data, the absolute mass, the
semi-major axis, the orbital period, and the mean density of the
transiting planets are determined. The transit method is particularly
suited for detecting ``very hot Jupiters'' (gas giant planets with an
orbital period in the range of 1 day ${\le}$ P ${\le}$ 3 days) and ``hot
Jupiters'' (3 days $<$ P ${\le}$ 9 days) \citep{key-36}.

The main difficulties for the transit method are two-fold; the geometrical
probability of a transit is small, and the photometric
accuracy required to detect the typical transit depth ($\sim 1$\%) is
demanding. This implies that wide-field accurate photometric
monitoring is necessary for successful detection of planets using the
transiting method. This is why only six extrasolar planets have been
discovered first by the transit method, and then confirmed by the
subsequent follow-up by the radial velocity measurement (3 additional
transiting planets were discovered first by the radial velocity method,
and their photometric transit signatures were found later).

Five out of the six confirmed transiting planets were discovered by the
Optical Gravitational Lensing Experiment (OGLE) team (\cite{key-2};
\cite{key-104}). Indeed they reported more than 130 extrasolar planet
candidates from about 150,000 stars which were observed with high
photometric precision of 0.015 mag or better. Follow-up observations using the
radial velocity method have so far confirmed three very hot Jupiters and
two hot Jupiters from the candidates (\cite{key-25}; \cite{key-26};
\cite{key-27}).  The single remaining transiting planet was discovered by
the Trans-Atlantic Exoplanet Survey network (TrEs) using small-aperture
(10cm), wide-field (6$^{\circ}$) CCD-based systems \citep{key-28}.

A number of transit search programs are currently on-going: the
Extrasolar Planet Occultation Research (EXPLORE) project \citep{key-29}
conducts deep transit searches of the Galactic plane using the MOSAIC
II camera at the CTIO 4 m telescope and the CFH12K camera at the 3.6 m
Canada-France-Hawaii Telescope. \citet{key-106} observed NGC 6940 using
the 2.5 m Isaac Newton Telescope with its Wide Field Camera, a mosaic
consisting of four 2048 ${\times}$ 4096 pixels CCDs. The MOA-I project observed 14 
Galactic Bulge fields using the 61 cm Boller and Chivens Telescope at the Mt.John 
University Observatory with three 2048 ${\times}$ 4096 pixels CCDs \citep{key-110}.  SuperWASP
\citep{key-55}, Vulcan \citep{key-56}, and others are also conducting very
wide-field transit surveys similar to TrEs. While transiting planet
candidates have been reported by these programs, the existence of the planets have 
not been confirmed yet. 

Transit search space missions are also planned, including COROT (French Space Agency
CNES, with participation of Austria, Belgium, Brazil, Germany, Spain,
ESP and ESTEC) and KEPLER (NASA). These missions are expected to achieve
a photometric precision up to 0.1\%, and aim at detecting terrestrial
planets in the habitable zone of their primary stars by the photometric transit method (\cite{key-102};
\cite{key-103}).

In this paper, we describe a prototype transit search using Subaru 
Suprime-Cam carried out in 2002. While our observing run covered only a small range in the
time domain (15 observing hours in total over 4 consecutive nights), the
field-of-view is significantly wider ($34^\prime\times27^\prime$), and
the survey magnitude is deeper than those of OGLE and EXPLORE.  Thus we
mainly look for single transit events, and perform a serious attempt to
test the feasibility of the photometric transiting planet search using
the Subaru Suprime-Cam for the first time.

The rest of the paper is organized as follows; section 2 describes our
observational strategies and data reduction procedures. In section 3,
we discuss the achieved photometric precision. Section 4 presents
candidate transiting objects; we found three single transit objects
(likely candidates for extrasolar planets) and two double transit
objects (possibly eclipsing binaries). Assuming circular orbits, 
we estimate the radii and the orbital
elements of the companions. Finally, section 5 is devoted to a summary and
further discussion. We also discovered eleven W UMa-like eclipsing
binary systems, which are reported in the Appendix.

\section{Observations and Data Reduction}

\subsection{Transit Search \label{subsec:transitsearch}}

In order to maximize the number of available stars, we selected a region
of $34'\times27'$ at the Galactic plane centered at $l=90^{\circ}$,
$b=0^{\circ}$ for our photometric survey.  We call it the Extrasolar
Planet-field (ESP-field).  The field was imaged with Suprime-Cam on the
8.2m Subaru Telescope which consists of 5 $\times$2 CCD chips with
2024$\times$4048 pixels each. Thus the detector scale corresponds to
$0^{''}.202$ per pixel.  We repeated photometric observations of the field with 60
second exposures and read-out times of 60 seconds each for 6 hours on
September 28 and 29, 2002, and for 1.5 hours on October 1 and 2, 2002.

The fractional transit depth ${\Delta}F$ is expressed as
%%%%%%%%%%%%%%%%%%%%%%%%%%%%%%%%%%%%%%%%%%%%%%%%%%%
\begin{equation}
\label{eq:deltaF}
 {\Delta}F=\left(\frac{R_{P}}{R_{*}}\right)^{2},
\end{equation}
%%%%%%%%%%%%%%%%%%%%%%%%%%%%%%%%%%%%%%%%%%%%%%%%%%%
where $R_{p}$ is the radius of a planet and $R_{*}$ is the radius of a
parent star; the depth is approximately 1 $\%$ for a Jupiter-size planet
orbiting around a Sun-like star.  Equation (\ref{eq:deltaF}) indicates
that relatively small-size stars, such as late-type dwarfs, are better
suited for transit detection. Since late-type dwarfs are brighter at
longer wavelength, we chose i'-band for the transit photometry; z'-band
is expected to be seriously compromised by high sky noise, and also the density
of stars in the z'-band image is too high (confusion limit).  In order to avoid the
effects of any systematic errors in flat fielding, we did not conduct dithering
 and tried to observe stars at constant position on the CCD. We obtained 
383 frames in the i' band for the entire observing
run. Excluding seven frames which suffered from the telescope tracking
errors, we used 376 images to search for transit events.
The sky was clear on September 28, October 1 and 2, but on September 29
there were occasionally thin clouds. The seeing was approximately
0.5-0.8 arcsec during the observing runs. We use GD 248 as a standard
star \citep{key-50} to determine the zero point of the $i'$-band
photometry.

Bias was estimated from the overscan region and subtracted from each
frame.  In carrying out this standard procedure, we used IRAF package
\footnote[1]{IRAF is distributed by the National Optical Astronomy
Observatories, which are operated by the Association of Universities for
Research in Astronomy, Inc., in cooperation with the National Science
Foundation.} for i'-band images and the Suprime-Cam Deep Field REDuction
(SDFRED) package (\cite{key-33}; \cite{key-34}) for multicolor
photometry. The dark current is completely negligible for 60 second exposures
\citep{key-39}.  The median of normalized dome frames was used as the
flat frame. An example of the reduced image of chip 2 for the $i'$ band
is shown in Figure 1. This image was obtained on September 29 and the
field-of-view is 1/10 of the total (10 chips).

%%%%%%%%%%%%%%%%%%%%%%%%%%%%%%%%%%%%%%%%%%%%%%%%%%%%%%%%%%%%%%%%%%%%%%
\begin{figure}
  \begin{center}
    \FigureFile(68mm,135mm){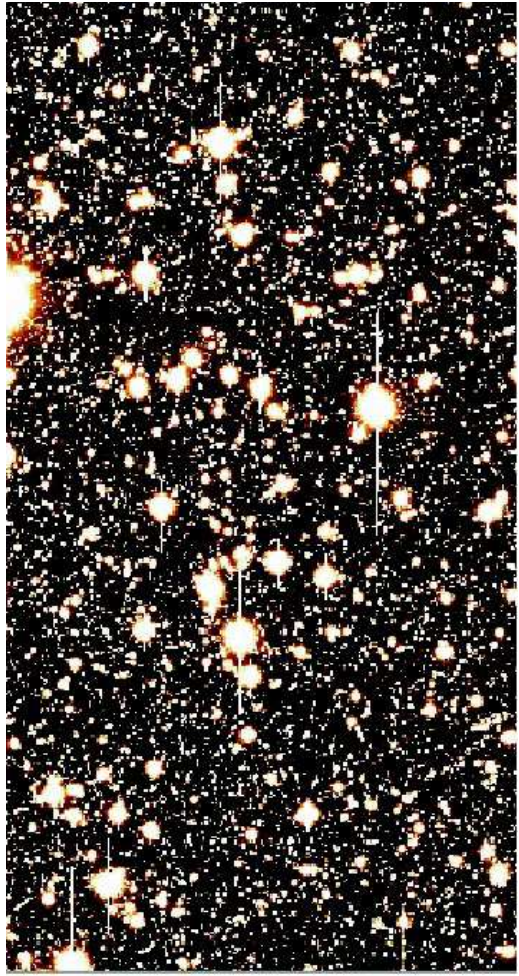}
  \end{center}
\caption{An example of the reduced $i'$-band image for chip 2.  The
field of view of this image approximately 7'$\times$13'. The entire
field of view of Suprime-Cam consists of ten chips (each chip has 2048
$\times$ 4096 pixels), and thus ten times as large as this image.  The
entire observed region (Esp-field) is centered at the Galactic plane of
$l=90^{\circ}$, $b=0^{\circ}$. }  \label{fig:chip2image}
\end{figure}
%%%%%%%%%%%%%%%%%%%%%%%%%%%%%%%%%%%%%%%%%%%%%%%%%%%%%%%%%%%%%%%%%%%%%%

We carried out aperture photometry using the IRAF package. The number of
identified point sources exceeds 100,000. We selected a value of 10
pixels for the aperture radius (i.e., approximately 3 times of FWHM for
the PSF). Those objects with their i'-band magnitudes satisfying
18${\le}$$i'$${\le}$23 were analyzed; we excluded i'$\le$18 objects
because of flux saturation, and i'$\ge$ 23 objects because of
insufficient counts for the accurate photometry.

Change in atmospheric airmass is the primary source of absolute
fluctuation in the light curves. In order to determine the atmospheric
change, we select more than 200 stars per CCD chip image as
references. We calculated the weighted mean fluxes of these reference
stars weighted by the flux error, which are represented by Poisson noise 
and sky noise for each star. While short-period variable stars might be 
included in the reference stars, they are very rare and their contribution 
is expected to be negligible; the fractional ratio of W UMa-like eclipsing variables, 
whose periods typically range between 0.2 and 1.0 day, is approximately one in 250-300 
main sequences. After correcting for the atmospheric airmass and transparency, we compute the 
fluxes of each star for every frame, and define the photometric precision 
for each star as the standard deviation among all the frames on September 
28 and 29. More specifically the photometric precision is given as
%%%%%%%%%%%%%%%%%%%%%%%%%%%%%%%%%%%%%%%%%%%%%%%%%%%%%%%%%%%%%%%%%%%%%%%
\begin{equation}
\sigma=\sqrt{\frac{1}{n} {\sum_{i=1}^n (F_{i}-\bar{F})}^{2}}.
\end{equation}
%%%%%%%%%%%%%%%%%%%%%%%%%%%%%%%%%%%%%%%%%%%%%%%%%%%%%%%%%%%%%%%%%%%%%%%
Here, ${\sigma}$ is the photometric precision and, $n$ is the number of frames
 for each night and, $F_{i}$ and $\bar{F}$ are the calibrated fluxes of each
 star for every frame and the mean fluxes of each star for every night,
 respectively.

\subsection{Multicolor Photometry}

We conducted multicolor photometry in the $B$, $R_{c}$, $z'$ bands to
determine spectral types of the stars. We obtained one image in each
band. We did not observe reference stars for these bands (except for
i'). Instead we used the mean flux of reference stars in the Subaru/XMM-Newton
Deep Survey Field (SXDS-field) observed on the same nights as our
observing run (Sekiguchi et al. 2006, in preparation) in order to
determine the zero point in the $B$, $R_{c}$ and $z'$ bands.  From the
inferred spectral type of each star, we estimate its radius and mass
approximately.  Moreover, we infer the distance of objects through 
the apparent magnitude and the typical absolute
magnitude for the spectral type.

\begin{figure}
  \begin{center}
\FigureFile(140mm,140mm){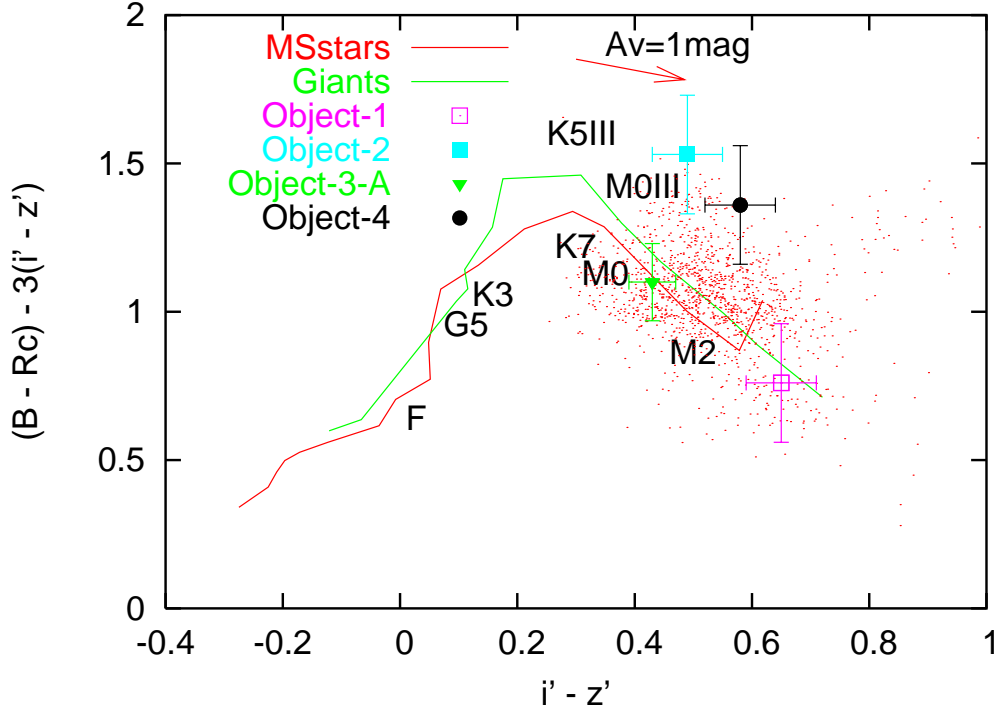}
  \end{center}
\caption{ Color-Color diagram for objects on chip 4 and transiting candidates
 discussed in \S 4. We plot only good photometry objects whose magnitude error is less 
than 0.055 magnitude and 0.2 magnitude for $i'-z'$ and $B-R_{c}-3(i'-z')$,
 respectively. The red 
  and green lines indicate the positions of the unreddened
  main sequences and giants, respectively (\cite{key-40};
  \cite{key-41}). The arrow represents the reddening
  vector.}\label{fig:color-color-plots}\
\end{figure}

Figure \ref{fig:color-color-plots} shows the color-color diagram for
detected stars in chip 4 and our transiting candidates in \S 4. In the figure,
 the red and green 
lines correspond to the positions of the unreddened main sequence and
giant stars, respectively. 
The arrow in Figure
\ref{fig:color-color-plots} is the reddening vector. We set the y-axis
as $B-R_{c}-3(i'-z')$ in order to clarify the locations of different
spectral types just for illustrative purposes.  
Figure \ref{fig:color-color-plots} indicates
that most of the monitored stars are of FGKM types, but that their
luminosity classes are difficult to infer from the multicolor
photometry alone.

%%%%%%%%%%%%%%%%%%%%%%%%%%%%%%%%%%%%%%%%%%%%%%%%%%%%%%%%%%%%%%%%%%%%%%
%%%%%%%%%%%%%%%%%%%%%%%%%%%%%%%%%%%%%%%%%%%%%%%%%%%%%%%%%%%%%%%%%%%%%%

\section{Photometry}

\subsection{Photometric Precision}

\begin{figure}
  \begin{center}
\begin{tabular}{cc}
\FigureFile(80mm,80mm){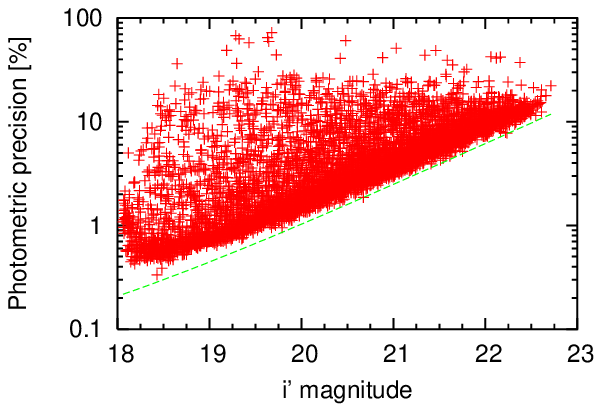}&\FigureFile(80mm,80mm){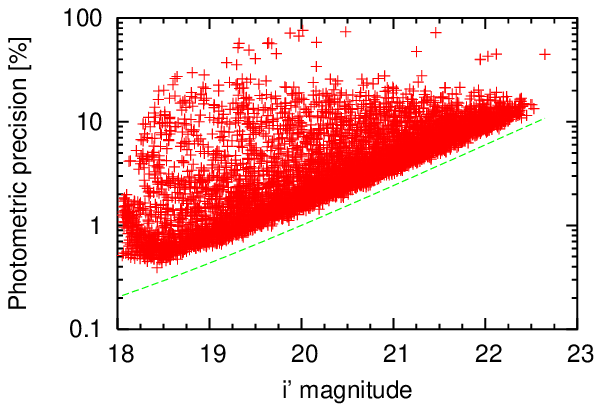}\\
\end{tabular}
  \end{center}
  \caption{Magnitude vs photometric precision for objects detected in
chip 2 on September 28 ($Left$) and September 29 ($Right$).
\label{fig:photo-precision}}
\end{figure}

Figure \ref{fig:photo-precision} shows the photometric precision for each star
(defined in \S \ref{subsec:transitsearch}) on chip 2 as a function
of its i'-band magnitude.  We note that all ten CCD chips show similar
photometric precisions.  The green line indicates the theoretical lower
bound, the Poisson noise of the stellar fluxes
and the typical sky noise in the imaging data on September 28.  Clearly
the photometric precision of most stars is close to the theoretical limit,
and a photometric precision as small as 0.4\% is achieved for stars
of i' $\sim$ 18. 
Stars of i'${\le}20$ lie significantly above a photometric precision of 10\% due to the contamination of the brightness of close saturated stars.

Figure \ref{fig:photo-precision} also suggests that the photometric precision
on September 29 is not better than that on September 28 due to the occasionally
thin clouds. However, since the influence of a little bad weather on September 29 is a small, the photometric precision on September 29 is enough to detect extrasolar planets.
 Among more than 100,000 stars within
the field of view, 27,000 objects have a photometric precision less than
3.0\% for each 60 second exposure.

Tabel 1 shows the resulting number of stars which achieve a given
photometric precision. For reference, very hot Jupiters and hot Jupiters
discovered by the OGLE and TrEs exhibit a flux decrease less than
3\%. Moreover, \citet{key-36} estimated that the ratio of main sequence
FGK stars which accompany transiting very hot Jupiters is approximately one 
in 3300-6700.  Therefore the photometric precision and
the number of stars in this prototype survey are sufficient for 
detecting very hot Jupiter candidates.
%%%%%%%%%%%%%%%%%%%%%%%%%%%%%%%%%%%%%%%%%%%%%%%%%%%%%%%%%%%%%%%%%%%%
%%%%%%%%%%%%%%%%%%%%%%%%%%%%%%%%%%%%%%%%%%%%%%%%%%%%%%%%%%%%%%%%%%%%

%%%%%%%%%%%%%%%%%%%%%%%%%%%%%%%%%%%%%%%%%%%%%%%%%%%%%%%%%%%%%%%%%%%%
\begin{longtable}{ccccccc}
  \caption{Summary of statistics for mean photometric precision achieved
in the September 28 and 29 run.}
\label{LT1} \hline\hline\ Photometric
precision& ${\le}5.0{\%}$ & ${\le}3.0{\%}$ & ${\le}2.0{\%}$&
${\le}1.5{\%}$&${\le}1.0{\%}$ & ${\le}0.50{\%}$ \\ \hline \endhead
Number of stars & 41,000 & 27,000 & 18,000 &13,500 & 7,700 & 1,000 \\
\hline\hline
\end{longtable} 
%%%%%%%%%%%%%%%%%%%%%%%%%%%%%%%%%%%%%%%%%%%%%%%%%%%%%%%%%%%%%%%%%%%%

\subsection{Selection of Candidates}

Typical transit durations of the previously detected systems (\cite{key-2};
\cite{key-104}; \cite{key-28}) is a few hours. Thus we define transit
candidates as light curves that have more than continuous fifteen
data points (approximately continuous 30 minutes data points) which
exhibit more than a 1${\sigma}$ flux decrement with respect to the average
flux during the observing run.
According to the selection criteria, we are left with around 1,000
objects out of more than 100,000 objects. We then inspected the light curves all selected objects by eye. Independently, in order to
detect periodic variable stars and double transit events, we conducted 
period analys using the phase dispersion minimization (PDM) 
algorithm described by 
\citet{key-52}. This algorithm minimizes scatter for a 
given data set phased to a series of test periods. The likelihood of
a given test period is given by the parameter ${\Theta}$. The minimum value
for ${\Theta}$ represents the minimum dispersion of data points for a given 
period. We selected around 2,000 objects with ${\Theta}$ less than 0.5, 
from more than 100,000 objects, as candidates for double transit objects and 
variable stars. We consider objects with full double decrements in 
stellar light as double transit objects. However, the folded light curves for 
double transit objects did not cover the full phase.
Next, we selected objects with sine-like light curves as 
W UMa eclipsing binary candidates.
Although the period of those eclipsing binary candidates are estimated 
by the PDM method, the true period might actually be commensurate with  
the estimated period. For this reason, the 
Lomb periodogram method (\cite{key-107}; 
\cite{key-51}) is employed for W UMa eclipsing binary candidates. 
The spectral power
obtained by this method gives the statistically most significant period.
Finally, we determine the W UMa eclipsing binary candidates according to the
following criteria, (1) the fluxes were not contaminated by nearby stars, 
(2) the objects were detected with all bands, and (3) the data points covered 
approximately the full phase.

\subsection{Estimation of Orbital Elements}

Detections of multiple photometric transit events and extensive radial
velocity follow-up are required to accurately determine orbital elements
of a companion around a parent star. Approximate values, however,
may be estimated even from the light curve of a single transit event,
assuming that (1) the parent star is dwarf, (2) $M_{*}$ and $R_{*}$
($M_{*}$ is the mass of the parent star and $R_{*}$ is the radius of the
parent star) are inferred from the spectral type (for instance,
\cite{key-38}), (3) the companion has a circular orbit, and (4) the
mass of the companion satisfies $M_{p}{\ll}M_{*}$ (see, for instance,
\cite{key-35}).
Then the orbital period $P$, the semi-major axis $a$, and the
inclination of the orbital plane with respect to the line of sight can be
explicitly written as
%%%%%%%%%%%%%%%%%%%%%%%%%%%%%%%%%%%%%%%%%%%%%%%%%%%%%%%%%%%%%%%%%%%%%%%%%
\begin{eqnarray}
\label{eq:P}
P&=&\frac{M_{*}}{R_{*}^3}
\frac{G{\pi}}{32}
\frac{(t_{T}^2-t_{F}^2)^{\frac{3}{2}}}{{({\Delta}F)^{\frac{3}{4}}}}, \\
\label{eq:a}
a&=&\left(\frac{P^{2}GM_{*}}{4{\pi}^{2}}\right)^{\frac{1}{3}}, \\
\label{eq:i}
{\cos}i &=& \frac{R_{*}}{a}\left\{\frac{(1-\sqrt{{\Delta}F})^{2}-[\sin^{2}(t_{F}{\pi}/P)/\sin^{2}(t_{T}{\pi}/P)](1+\sqrt{{\Delta}F})^{2}}{1-[\sin^{2}(t_{F}{\pi}/P)/\sin^{2}(t_{T}{\pi}/P)]}\right\}^\frac{1}{2},
\end{eqnarray}
%%%%%%%%%%%%%%%%%%%%%%%%%%%%%%%%%%%%%%%%%%%%%%%%%%%%%%%%%%%%%%%%%%%%%%%%%
where $G$ is the gravitational constant, and $t_{T}$ and $t_{F}$ denote
the total transit duration and the ``flat-bottom'' duration, respectively.

We attempted to compute the orbital elements of OGLE and TrEs planets
using the above equations. The result is shown in Table \ref{orbit}. 
We found that while the estimated periods are different from the actual 
ones by several hours, the semi-major axes and the inclinations agree 
within around 0.01 AU and a few degrees, respectively. 

%%%%%%%%%%%%%%%%%%%%%%%%%%%%%%%%%%%%%%%%%%%%%%%%%%%%%%%%%%%%%%%%%%%%
\begin{longtable}{ccccccc}
  \caption{Orbital elemensts for extrasolar planets and their estimated orbital elements using equations (\ref{eq:deltaF}) and (\ref{eq:P}) to (\ref{eq:i}).}\label{orbit}
 \hline\hline\
Object & P  & a & i & ${\Delta}F$ & t${_T}$& t${_F}$\\
~& (observed ${\vert}$ estimated ) & (observed ${\vert}$ estimated)& (observed ${\vert}$ estimated)  &~&~&~\\
~& $[day]$& $[AU]$ & [$^{\circ}$] & [\%] & [min] & [min]\\
\hline
\endhead
\hline
\endfoot
\hline
\multicolumn{3}{l}{\hbox to 0pt{\parbox{180mm}{\footnotesize
\footnotemark[${a}$] \cite{key-105}, \footnotemark[${b}$] \cite{key-108}, \footnotemark[${c}$] \cite{key-27},
\footnotemark[${d}$] \cite{key-109}, \footnotemark[${e}$] \cite{key-28} \\
 }}}
\endlastfoot
OGLE-TR-10b$^{a}$ &3.101386~ ${\vert}$ 2.3 & 0.04162 ${\vert}$ 0.034 & 89.2~~~ ${\vert}$ 86 & 1.61& 174 & 141\\

OGLE-TR-56b$^{b}$ & 1.2119189${\vert}$ 1.3 & 0.0225~ ${\vert}$ 0.024 & 81~~~~~ ${\vert}$ 81  & 1.32& 113& 71 \\
OGLE-TR-111b$^{c}$ & 4.01610~~ ${\vert}$ 3.8 & 0.047~~ ${\vert}$ 0.044 & 86.5-90 ${\vert}$ 88 &1.46& 164 & 124 \\
OGLE-TR-113b$^{d}$ & 1.43250~~ ${\vert}$ 1.5 & 0.0228~ ${\vert}$ 0.023 & 85-90~~${\vert}$ 87 &2.10 &  114& 82 \\
OGLE-TR-132b$^{d}$ & 1.68965~~ ${\vert}$ 2.4 & 0.0306~ ${\vert}$ 0.039 & 78-90~~ ${\vert}$ 84 &0.70& 165 & 125\\
TrES-1$^{e}$ & 3.030065~ ${\vert}$ 3.1 & 0.0393~ ${\vert}$ 0.039 & 88.5~~~ ${\vert}$ 89 & 1.69 & 154 & 116\\

\hline\hline
\end{longtable} 
%%%%%%%%%%%%%%%%%%%%%%%%%%%%%%%%%%%%%%%%%%%%%%%%%%%%%%%%%%%%%%%%%%%%
In order to derive the transit depth, the transit duration and the
flat-bottom duration, we need a template light curve for the fit.  For
that purpose, we use the perturbation result derived by
\citet{key-100}. The template flux is given as 
\begin{equation}
 F=1-\frac{f(\gamma,\rho)}{\pi(1-u_1/3-u_2/6)}.
\end{equation}
Here, $u_1$, $u_2$ are limb darkening parameters and they are defined as
\begin{equation}
I(\mu)/I(0)=\left[1-u_1(1-\mu)-u_2(1-\mu)^2\right].
\end{equation}
Here, $\mu$ is an angle between the star's radius vector and the line of sight. Moreover, $f(\gamma,\rho)$ is given as 
\begin{equation}
 f(\gamma,\rho)=\cases{
  \pi\gamma^2\left[1-u_1-u_2\left(2-\rho^2-\frac{1}{2}\gamma^2\right)+(u_1+2u_2)W_1\right] & $\rho<1-\gamma$\cr
  \left(1-u_1-\frac{3}{2}u_2\right)\left[\gamma^2\cos^{-1}\left(\frac{\zeta}{\gamma}\right)+\sin^{-1}z_0-\rho z_0\right]\cr
  +\frac{1}{2}u_2\rho\left[(\rho+2\zeta)\gamma^2\cos^{-1}\left(\frac{\zeta}{\gamma}\right)-z_0\left(\rho\zeta+2\gamma^2\right)\right]\cr
+(u_1+2u_2)W_3 & $1-\gamma<\rho<1+\gamma$\cr
  0 & $\rho>1+\gamma$,\cr
  }
\end{equation}
where $\gamma$, $\rho$, $\eta_p$, $z_0$, $\zeta$ are defined as 
\begin{equation}
\gamma=R_p/R_*,
\end{equation}
\begin{equation}  
\rho=1+\eta_p=\frac{\sqrt{X_p^2+Z_p^2}}{R_*}
\end{equation}
\begin{equation}
 z_0=\frac{\sqrt{(\gamma^2-\eta_p^2)\left[(\eta_p+2)^2-\gamma^2\right]}}{2(1+\eta_p)}
\end{equation}
\begin{equation}
 \zeta=\frac{2\eta_p+\gamma^2+\eta_p^2}{2(1+\eta_p)}.
\end{equation}
Here, $X_p$ and $Z_p$ are the positions of the center of the planet in the star's central coordinate system. $W_1$ and $W_3$ are integrals which are defined in the Appendix of \citet{key-100}.
 For simplicity, we do not consider the effect of limb darkening($u_1=u_2=0$). After
the fit, we obtain the orbital elements by substituting the values of
$\Delta F$, $t_{\rm F}$, and $t_{\rm T}$ into equations
(\ref{eq:deltaF}) and (\ref{eq:P}) to (\ref{eq:i}).

%%%%%%%%%%%%%%%%%%%%%%%%%%%%%%%%%%%%%%%%%%%%%%%%%%%%%%%%%%%%%%%%%%%%%%%%%
\begin{longtable}{ccccccc}
  \caption{Observed parameters for the transiting candidates.}
\label{LT2}
 \hline\hline\
Object & RA & Dec & B [mag]& R$_{c}$ [mag]& i' [AB mag]& z' [AB mag] \\
\hline
\endhead
Object-1 &$\timeform{21h13m35s.7}$& $\timeform{48D15'20".6}$ & 21.55${\pm}$0.02 & 18.84${\pm}$0.04 & 18.34${\pm}$0.05 & 17.69${\pm}$0.04\\

Object-2 & $\timeform{21h13m37s.2}$& $\timeform{48D10'37".8}$ &22.04${\pm}$0.02 & 19.04${\pm}$0.03 & 18.50${\pm}$0.05& 18.01${\pm}$0.04  \\
Object-3-A& $\timeform{21h11m42s.5}$& $\timeform{48D9'46".4}$ & 21.13${\pm}$0.02 & 18.74${\pm}$0.02 & 18.47${\pm}$0.03 & 18.04${\pm}$0.03 \\
Object-3-B & $\timeform{21h11m42s.5}$& $\timeform{48D9'47".0}$ & 23.32${\pm}$0.09& 20.10${\pm}$0.03 & 19.40${\pm}$0.03 &18.71${\pm}$0.03\\

Object-4 &$\timeform{21h13m17s.7}$& $\timeform{48D6'43".7}$ & 23.73${\pm}$0.06& 20.63${\pm}$0.03 & 19.86${\pm}$0.05 & 19.28${\pm}$0.04\\
Object-5 &$\timeform{21h13m12s.0}$& $\timeform{48D18'0".6}$ & --- & 21.89${\pm}$0.07 & 20.53${\pm}$0.05 & 19.81${\pm}$0.06\\
\hline\hline
\end{longtable} 
%%%%%%%%%%%%%%%%%%%%%%%%%%%%%%%%%%%%%%%%%%%%%%%%%%%%%%%%%%%%%%%%%%%%%%%%%

%%%%%%%%%%%%%%%%%%%%%%%%%%%%%%%%%%%%%%%%%%%%%%%%%%%%%%%%%%%%%%%%%%%%%%%%%
\begin{longtable}{ccccccccccc}
  \caption{Estimated parameters for the
  transiting candidates assuming they are dwarfs.
 Total transit duration and the flat-bottom
  duration are represented by $t_{T}$ and $t_{F}$,
  respectively.}\label{LT3}
 \hline\hline\
Object & SpT& A$_V$&${\Delta}F$& t$_{T}$ & t$_{F}$& P& a & i & Dist & Radi\\
~& ~&[mag] & $ [{\%}] $& $[min]$ & $[min]$& $[day]$& $[AU]$ & [$^{\circ}$] & $[kpc]$ & [$R_{Jup}$]\\
\hline
\endhead
Object-1 & G0V/K5V & 3.1 & 3.9${\pm}$0.8 &76-92 & 0-34 &0.3-1.8&0.01-0.02&67-84 & 1.0-3.0 & 1.4-2.2\\
~& M0V/M5V & 1.1 & 4.4${\pm}$0.8& 91-111& 0&2.1-6.9 & 0.02-0.03 &86-89 & 0.3-1.2 &0.6-1.3 \\
Object-2 & K5V/M0V &1.1 &4.9${\pm}$0.5&101-106& 23-34 &1.5-2.5& 0.02-0.03 &84-86 & 1.3-2.3 &1.3-1.6 \\
Object-3-A& K0V/M2V &1.4 &3.2${\pm}$0.7 & 106-111 &0-39&1.7-5.6&0.03-0.05& 83-88 & 1.2-3.0\ & 0.9-1.5\\

Object-4 & K5V/M0V & 1.6 &17.0${\pm}$1.6 & 86-93 & 0 &1.018 & 0.02 & 84 & 2.3-4.0 &2.5-3.0\\
Object-5 & ---&---&48.2${\pm}$4.2& 90 & 0 & 0.942 & --- &---- & ---& ---\\

\hline\hline
\end{longtable} 
%%%%%%%%%%%%%%%%%%%%%%%%%%%%%%%%%%%%%%%%%%%%%%%%%%%%%%%%%%%%%%%%%%%%%%%%%

%%%%%%%%%%%%%%%%%%%%%%%%%%%%%%%%%%%%%%%%%%%%%%%%%%%%%%%%%%%%%%%%%%%%%%%%%
 \begin{figure}
  \begin{center}
\begin{tabular}{cc}
\FigureFile(80mm,80mm){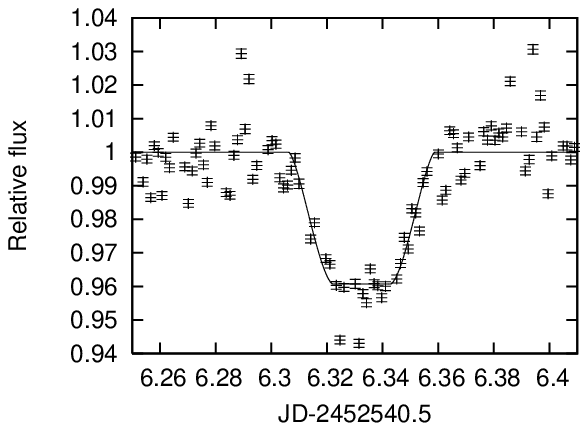}&\FigureFile(60mm,60mm){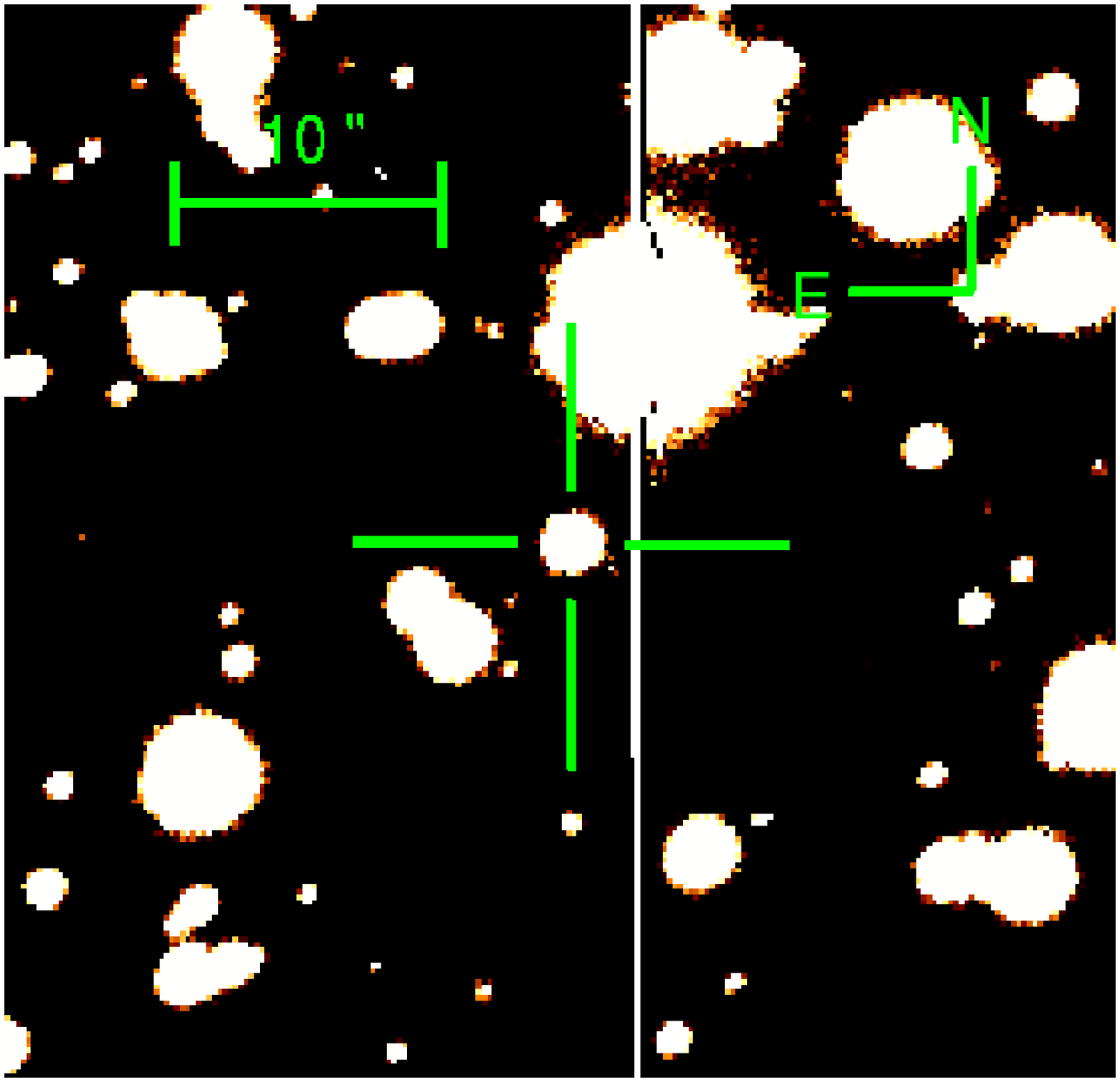}\\
~~~~~Object-1& ~~~~~Object-1 (Image)\\

\FigureFile(80mm,80mm){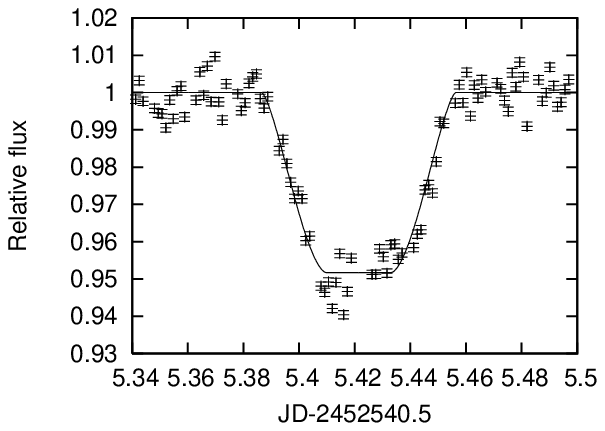} & \FigureFile(60mm,60mm){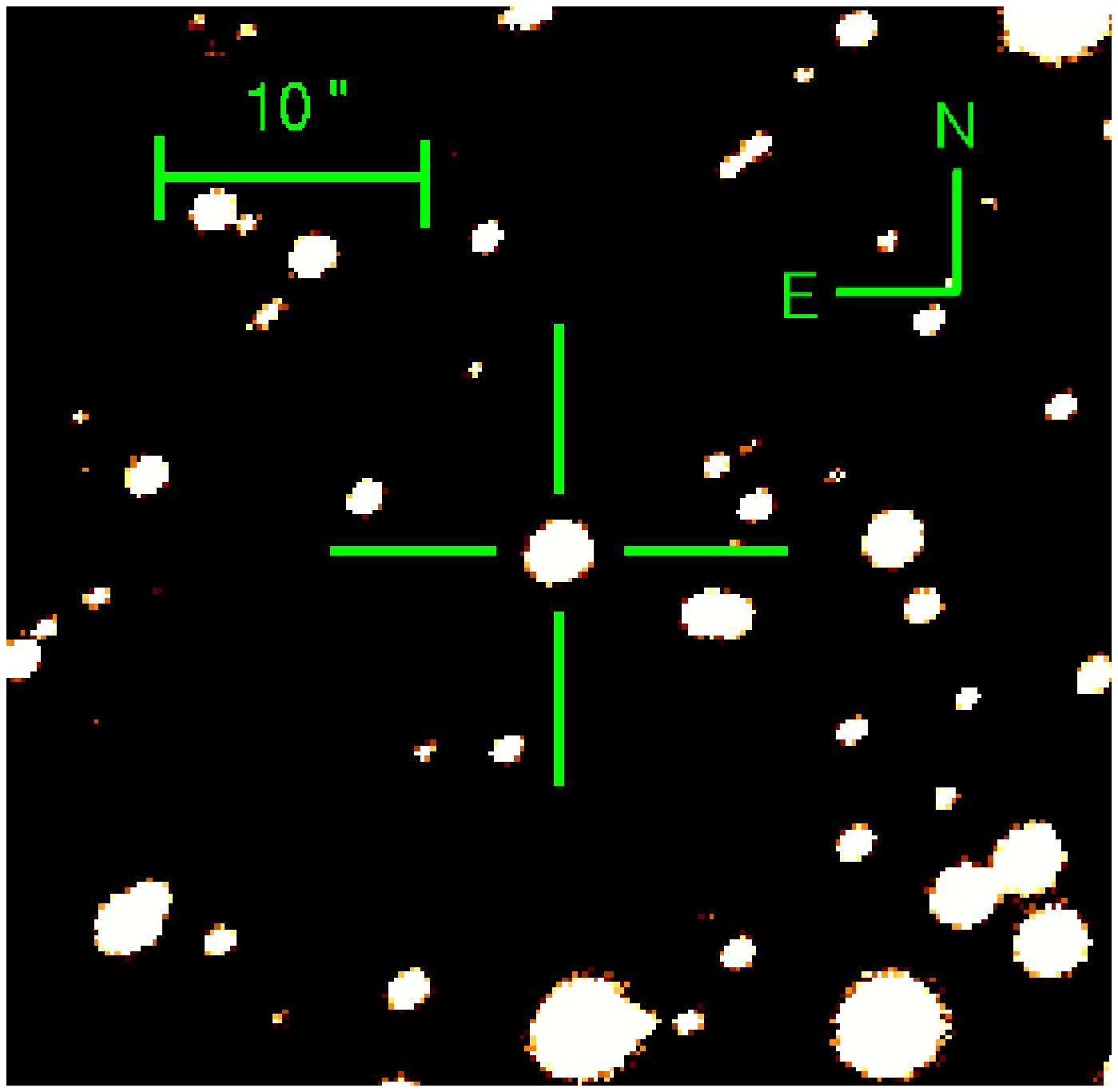}\\
~~~~~Object-2& Object-2 (Image)\\

\FigureFile(80mm,80mm){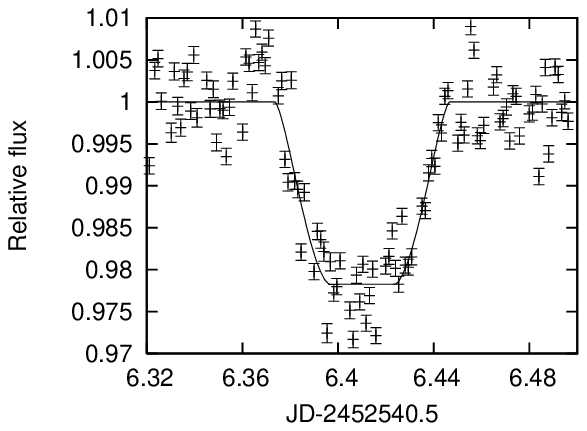} & \FigureFile(60mm,60mm){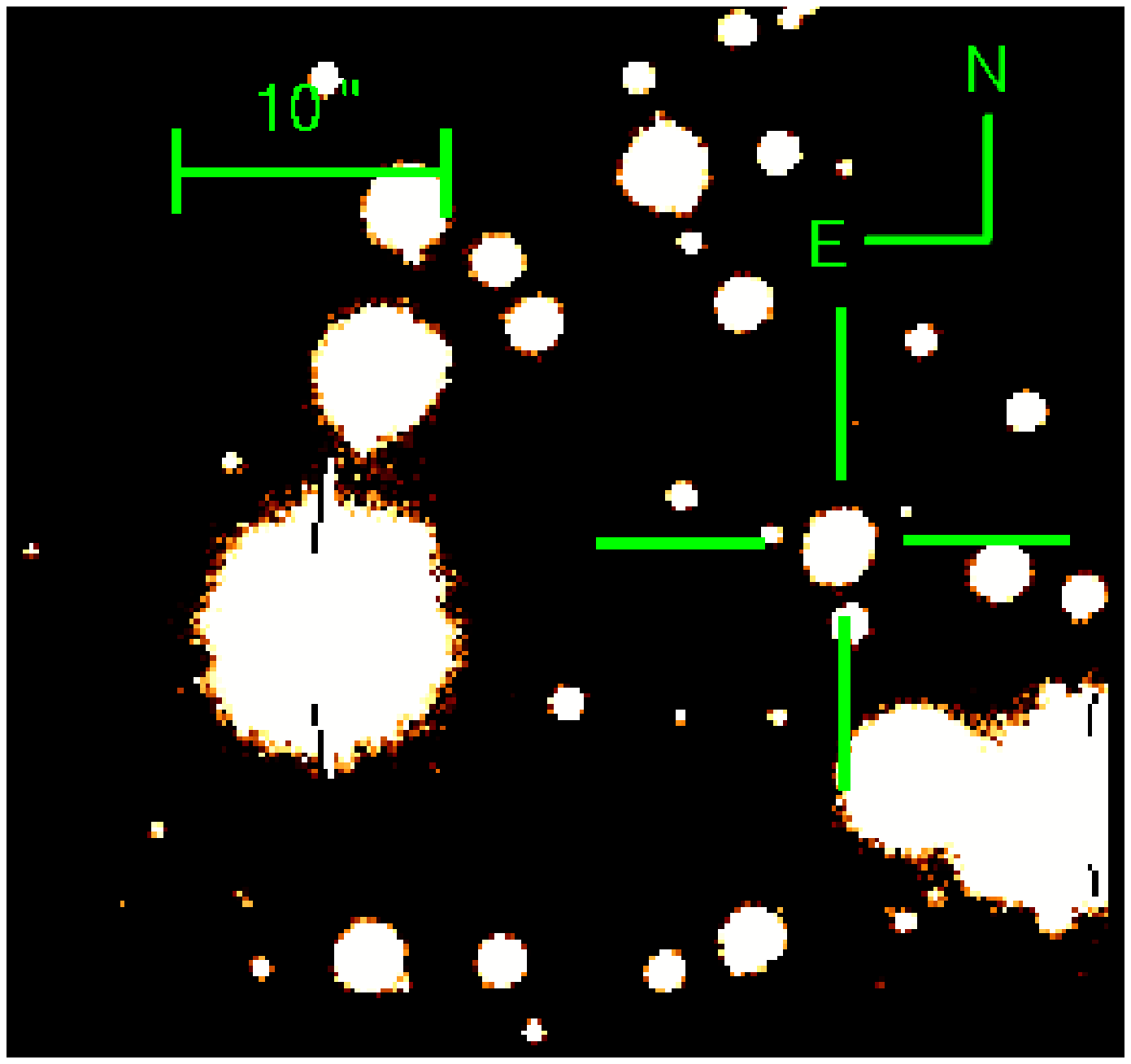}\\
~~~~~Object-3& Object-3 (Image)\\

\end{tabular}
  \end{center}
  \caption{$Left$: Light curves for the single transiting candidates. 
The best-fit light curve models are based on \citet{key-100}. We
  assume a K0 dwarf for Object-1 and Object-3, and a K5 dwarf for Object-2. 
$Right$: Images for the single transiting candidates. Object-3 turns out to be 
a double star.}\label{fig:lightcurve}
\end{figure}
%%%%%%%%%%%%%%%%%%%%%%%%%%%%%%%%%%%%%%%%%%%%%%%%%%%%%%%%%%%%%%%%%%%%%%%%%
%%%%%%%%%%%%%%%%%%%%%%%%%%%%%%%%%%%%%%%%%%%%%%%%%%%%%%%%%%%%%%%%%%%%%%%%%
\begin{figure}
  \begin{center}
\begin{tabular}{cc}
\FigureFile(80mm,80mm){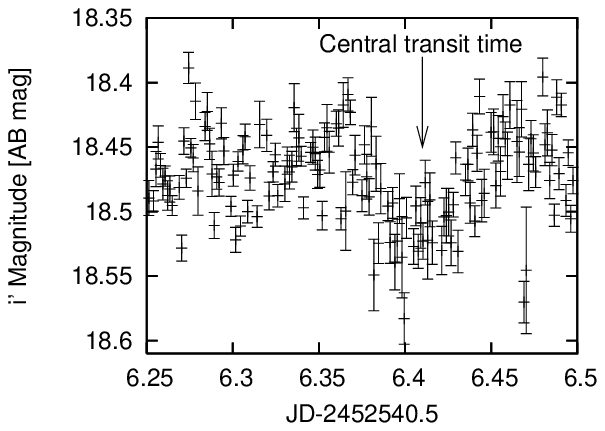}&\FigureFile(80mm,80mm){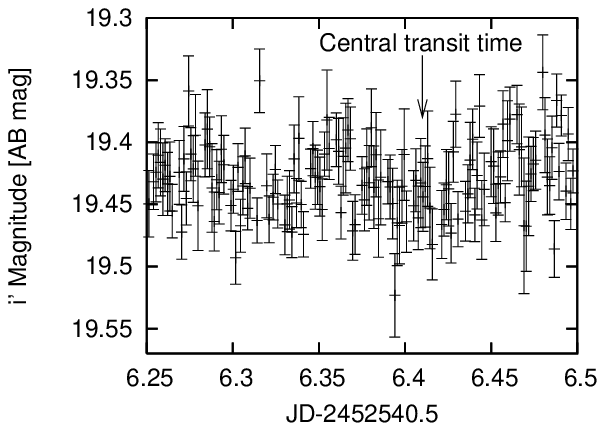}\\
\end{tabular}
  \end{center}
  \caption{Light curves of Object-3-A ({\it Left}) and
  Object-3-B ({\it Right}) from the PSF photometry. 
The arrows indicates the central transit time estimated from
Figure \ref{fig:lightcurve}.}\label{fig:obj3ab}
\end{figure}
%%%%%%%%%%%%%%%%%%%%%%%%%%%%%%%%%%%%%%%%%%%%%%%%%%%%%%%%%%%%%%%%%%%%%%%%%

%%%%%%%%%%%%%%%%%%%%%%%%%%%%%%%%%%%%%%%%%%%%%%%%%%%%%%%%%%%%%%%%%%%%%%%%%
 \begin{figure}
  \begin{center}
\begin{tabular}{cc}
\FigureFile(80mm,80mm){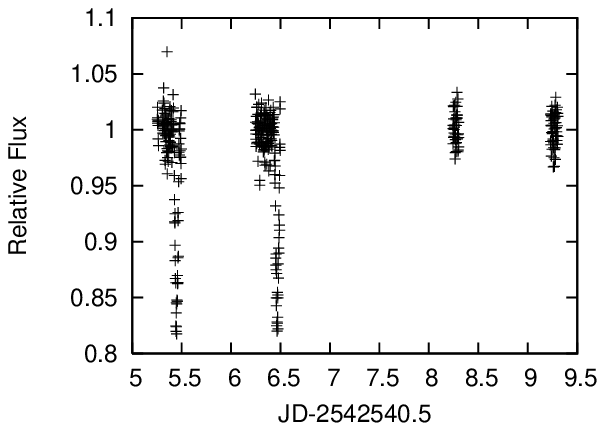}&\FigureFile(80mm,80mm){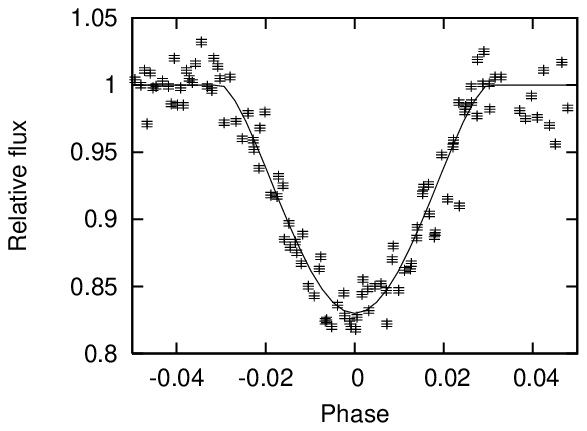}\\
~~~~~Object-4 & ~~~~~Object-4 (Folded light curve)\\
\FigureFile(80mm,80mm){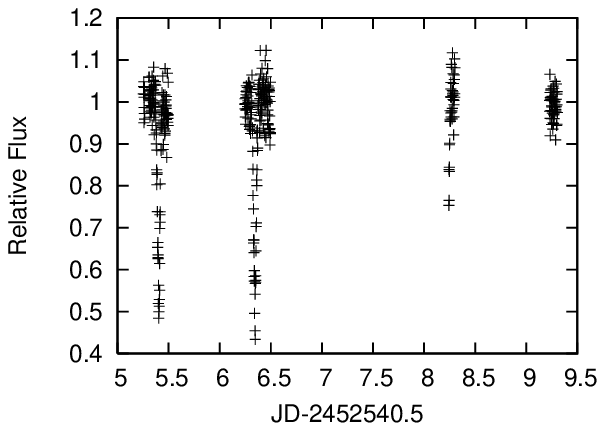}&\FigureFile(80mm,80mm){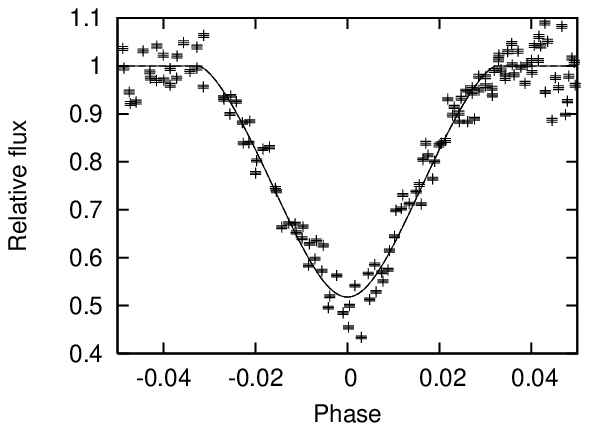}\\
~~~~~Object-5 & ~~~~~Object-5 (Folded light curve)\\
\end{tabular}
  \end{center}
  \caption{$Left$: Light curves for the double transiting candidates during this observing run. $Right$: Folded light curves. The best-fit light
  curve models are based on \citet{key-100}. We
  assume a K5 dwarf for Object-4. For Object-5 whose spectral type is unknown, we
  tentatively assume 1 ${\RO}$ for its parent star.}\label{fig:lightcurve2}
\end{figure}
%%%%%%%%%%%%%%%%%%%%%%%%%%%%%%%%%%%%%%%%%%%%%%%%%%%%%%%%%%%%%%%%%%%%%%%%%

\section{Candidate Transiting Objects}

After visual inspection of around 1000 objects which satisfied our
selection criteria, we found five strong transit candidates; three objects
(Object-1, Object-2 and Object-3-A) exhibit a single full transit-like
event. Their fractional flux decrements ($\sim 5$\%) for a couple of
hours are consistent with a planetary transit, and spectroscopic 
photometric
and/or photometric follow-up observations is justified.  The other two
objects (Object-4 and Object-5) show double transit events, and their
transit depths indicate that they are binary star systems. Further
information of those candidates is summarized in Tables \ref{LT2} and
\ref{LT3}, and their lightcurves are shown in Figure
\ref{fig:lightcurve} to Figure \ref{fig:lightcurve2}.  The error in the 
transit depth quoted in Table
\ref{LT3} corresponds to the standard deviation of the baseline flux of
the light curve during the out-of-transit phase.  We briefly discuss
each object below.

\subsection{Object 1}

Multicolor photometry indicates that Object-1 ($i'$=18.34 AB mag) is
either a G0/K5 or an M0/M5 star.  It is likely that Object-1 is a dwarf
because its distance exceeds the Galaxy scale ($> 17$kpc) if it is a
giant. Adopting typical parameters for a G0/K5 or M0/M5 dwarf, we estimate
the companion size and orbital elements using equations (\ref{eq:deltaF}) 
and (\ref{eq:P}) to (\ref{eq:i}) which are summarized in Table \ref{LT3}. 

Depending on the choice of the spectral type, and therefore the adopted
radius, of the parent star, the companion takes two different sets of
parameters. If it is an M dwarf and Av is derived from its observed
color and an unreddened standard M2 dwarf, the distance is estimated as
0.3-1.2 kpc. The companion radius is $(0.6-1.3)R_{\rm J}$, the orbital
period is 2.1-6.9 day, the semi-major axis is 0.02-0.03 AU, and the
inclination is 86-89$^{\circ}$. This set of parameter is consistent
with a Jupiter size companion.

\subsection{Object 2}

 Multicolor photometry indicates that Object-2 ($i'$=18.50 AB mag) is
likely to be a late K giant, but if it is a giant, the inferred
distance of 84 kpc exceeds the Galaxy's size. Moreover, given the observed
scatter of objects in Figure \ref{fig:color-color-plots}, a dwarf could
well be located at the position of Object-2 in the color-color diagram.
Therefore, it is quite possible that Object-2 is a K5/M0 dwarf. 
Adopting this interpretation, we obtain a distance of 1.3-2.3 kpc, 
a companion radius
of $(1.3-1.6)R_{\rm J}$, an orbital period of 1.5-2.5
days, a semi-major axis of 0.02-0.03 AU, and an inclination of
84-86$^{\circ}$.

\subsection{Object 3}

Object-3 turned out to be a very close double star with a separation of
$\sim$0.6 arcsec; their individual fluxes cannot be separated via
aperture photometry. So we reanalyzed Object-3 using PSF photometry. The
results are shown in Figure \ref{fig:obj3ab}. We find that Object-3 consists
of Object-3-A ($i'$ = 18.47 AB mag) and Object-3-B ($i'$ = 19.40 AB mag)
and the transit-like feature preferentially appears in Object-3-A.

Multicolor photometry suggests that Object-3-A is a K0/M2 star. We
estimate the depth, the transit durations and the flat-bottom durations
from the model light curves plotted in Figure \ref{fig:lightcurve} since
the lightcurve for Object-3-A from the PSF photometry is noisier and
still may not completely free from flux contamination of Object-3-B.
Ascribing the total flux decrement to a companion around Object-3-A, we
derive the transit depth of Object-3-A is 3.2{\%} using the mean
brightness ratio of Object-3-A and Object-3-B from the PSF photometry.

Assuming that a companion orbits around a K0/M2 dwarf, the distance is
1.3-2.3 kpc and the companion radius is $(0.9-1.5)R_{\rm J}$, the
orbital period is 1.7-5.6 days, the semi-major axis is 0.03-0.05 AU, and
the inclination is 83-88$^{\circ}$.

\subsection{Object 4}

Object-4 ($i'$=19.86 AB mag) exhibited double transits of a fairly large
depth of 17.0${\pm}$1.6{\%}.  The PDM method suggests an orbital period
of 1.018 days, and multicolor photometry indicates that Object-4 is a
K5/M0 star. The light curve has a transit duration of 86-93 minutes and
does not show a flat-bottomed shape. Assuming that Object-4 is a K5/M0
dwarf, the distance is 2.3-4.0 kpc. For a circular-orbiting companion,
the semi-major axis is 0.02 AU, the inclination is 84$^{\circ}$, the
companion radius is $(2.5-3.0)R_{\rm J}$. Thus the companion of Object-4
is likely to be a late M dwarf. Since the observing run did not cover
the full phase, we were unable to detect a secondary eclipse.  

\subsection{Object 5}

Object-5 ($i'$ = 20.53 AB mag) also exhibited double transits of a
significant depth of 48.2${\pm}$4.2{\%}. Because of the large depth,
Object-5 is likely to be an eclipsing binary system with an orbital
period of 0.942 days. Because Object-5 is very faint, we were not able
to identify Object-5 in our $B$-band data. Thus we cannot determine
its spectral type and the other parameters.

\section{Summary and Discussion}  
We have carried out a photometric search for transiting extrasolar
planets using Subaru Suprime-Cam. Out of about 100,000 stars monitored,
we obtain 7,700 (27,000) light curves with photometric precision below
1\% (3\%), required to detect extrasolar planets by the transit
method. This result thus demonstrates that Suprime-Cam indeed has the
photometric stability and accuracy required for a transiting planet
survey.

During this observing run, we detected three transiting planetary
candidates (i'-band magnitude around 18.5) which exhibit a single full
transit-like light curve with a depth of $<5\%$. Thus it is worthwhile
attempting spectroscopic and photometric follow-up observations of these
candidates. Adopting typical parameters for the parent stars from their
spectral type, we infer that their semi-major axes are less than 0.05
AU. If a Jupiter mass object orbits at a distance of 0.05 AU from a
central star of 1 ${\MO}$ star, the radial velocity amplitude exceeds
100 m/s. A faint (i'$\sim$ 18.5) object with 100 m/s radial 
velocity
would be detectable via a normal Tr-Ar wavelength calibration (i.e., without an iodine 
cell) with 7,200 second exposures per frame using the high-dispersion
spectrograph of a large telescope, such as Subaru, HDS
\citep{HDS,Sato02,Winn,Narita}. 

Finally we discuss a strategy of follow-up photometry and possible future 
transit surveys using Subaru. 
The predicted detection number $N_{p}$ of very hot Jupiters by a transit survey is roughly estimated as follows

%%%%%%%%%%%%%%%%%%%%%%%%%%%%%%%%%%%%%%%%%%%%%%%%%%%%%%%%%%%%%%%%%%%%%%%%%
\begin{equation}
 N_{p} {\sim} NF_{p}P_{vis}.
\end{equation}
%%%%%%%%%%%%%%%%%%%%%%%%%%%%%%%%%%%%%%%%%%%%%%%%%%%%%%%%%%%%%%%%%%%%%%%%%
Here, $N$ is the observed number of single FGKM dwarfs that have an
adequate photometric precision for very hot Jupiters detections, $F_{p}$
is the fraction of field stars that accompany transiting very hot
Jupiters,and $P_{vis}$ is the probability that at least double full
transits will occur during the observing run \citep{key-29}. In our 
observations, a photometric precision
within 3.0\% is achieved for approximately 27,000 objects. Assuming that 
most of them are dwarf and the binary fraction is 0.5, the expected value 
of $N$ is 13,500. $F_{p}$ is
one in 3300-6700 single main sequence FGK stars for very hot Jupiters according to
\citet{key-36}. If we conduct continuous 9 hours of observation each night 
for 10 days, $P_{vis}$ is 0.49 and $N_p$ is approximately 1.0-2.0. A longer 
observing run is needed to increase $P_{vis}$ for detection of multiple transits. Given the highly competitive time
allocation for Subaru, it is not realistic to look for the multiple
transit features of (very) hot Jupiters. Our current result showed,
however, that single transit candidates can be located with Suprime-Cam
relatively easily. Moreover, a future transit survey of a denser region of stars 
due to the good seeing at Subaru site would make it possible to detect more 
single transit candidates.
Then those candidates could be 
observed for follow-up
photometry with a smaller telescope. For instance, we plan to attempt
follow-up of our candidates (i'$\sim$ 18.5) with the Nayuta (2 m telescope 
of Nishiharima Astronomical Observatory, Japan) and the 1.5 m telescope of Gunma 
Astronomical Observatory. Since the transit depth of our candidates is 
approximately 4\%, we would be able to detect multiple transit features 
with 120-360 second exposures using these telescopes. Such complementary 
observations by a 2 m class telescope would make the transit 
survey using a very large telescope more effective and productive and lead to confirmed detection of extrasolar planets.

\bigskip

We thank Yasuhiro Ohta for providing us with a numerical routine to compute
the lightcurve of planetary transits.  This work is supported by a
Grant-in-Aid for JSPS Fellows (No.52791) and ``The 21st Century COE
Program of Origin and Evolution of Planetary System'' from the Ministry of
Education, Culture, Sports, and Technology.

%%%%%%%%%%%%%%%%%%%%%%%%%%%%%%%%%%%%%%%

\appendix

\section{W UMa-like eclipsing binary  candidates}

We also detected eleven W UMa-like objects whose period is less than 0.4
days. The light curves are shown in Figure 7. The observational results 
are shown in
Table 5 and Table 6. In order to determine the distance, we used the
color-magnitude relation and period-color relation \citep{key-53} for
the spectral type between early F and middle K stars. These are written
as
%%%%%%%%%%%%%%%%%%%%%%%%%%%%%%%%%%%%%%%%%%%%%%%%%%%%%%%%%%%%%%%%%%%%%%%%%
\begin{equation}
 M_{V}=-4.44{\log}P+3.02(B-V)+0.12,
\end{equation}
\begin{equation}
 (B-V)=0.04{\times}P^{-2.25}.
 \end{equation}
%%%%%%%%%%%%%%%%%%%%%%%%%%%%%%%%%%%%%%%%%%%%%%%%%%%%%%%%%%%%%%%%%%%%%%%%%
Here $M_{V}$ is the absolute magnitude and $P$ is the
period. Substituting equation (A2) for equation (A1), we can calculate the
absolute magnitude without observations for the $B$ band and $V$
band. If the spectral type of objects is not between early F and middle
K, the distance is estimated by the typical absolute magnitude for the
spectral type of the objects.

%%%%%%%%%%%%%%%%%%%%%%%%%%%%%%%%%%%%%%%%%%%%%%%%%%%%%%%%%%%%%%%%%%%%%%%%%
\begin{longtable}{ccccccc}
  \caption{Observed parameters for W UMa-like objects.}\label{tab:LT4}
 \hline\hline\
Object & RA &Dec & B [mag] & R$_{c}$ [mag] & i' [AB mag] & z' [AB mag] \\
~& ~&~&(Phase) & (Phase) & (Phase) & (Phase) \\
\hline
\endhead
Object-6& $\timeform{21h13m24s.1}$& $\timeform{48D29'49".9}$&23.20${\pm}$0.05 & 19.66${\pm}$0.03 & 18.78${\pm}$0.10 & 18.04${\pm}$0.04 \\
~& ~& ~& (0.61) & (0.79) & (0.78) & (0.82) \\
Object-7 & $\timeform{21h12m29s.2}$& $\timeform{48D7'31".8}$& 21.83${\pm}$0.03& 19.28${\pm}$0.03 & 19.08${\pm}$0.06 &18.65${\pm}$0.05   \\
~& ~& ~& (0.56) & (0.94) & (0.43) & (0.98) \\

Object-8 & $\timeform{21h11m4s.9}$& $\timeform{48D8'3".6}$ &22.54${\pm}$0.04 & 19.53${\pm}$0.03 & 18.95${\pm}$0.06 & 18.18${\pm}$0.05  \\
~& ~&~ & (0.35) & (0.22) & (0.61) & (0.26) \\

Object-9 &$\timeform{21h12m1s.0}$& $\timeform{48D17'29".2}$&24.33${\pm}$0.13 & 21.45${\pm}$0.04 & 20.48${\pm}$0.03& 19.83${\pm}$0.03  \\
~&~&~& (0.69) & (0.48) & (0.17) & (0.52) \\

Object-10 &$\timeform{21h10m33s.1}$& $\timeform{48D15'20".7}$ & 24.40${\pm}$0.09& 20.86${\pm}$0.03 & 20.38${\pm}$0.06 & 19.39${\pm}$0.05\\
~& ~&~& (0.76) & (0.61) & (0.64) & (0.88) \\

Object-11 &$\timeform{21h10m31s.9}$& $\timeform{48D8'18".5}$ & 23.91${\pm}$0.07 & 20.83${\pm}$0.04 & 20.21${\pm}$0.06 & 19.39${\pm}$0.05  \\
~&~&~& (0.95) & (0.70) & (0.73) & (0.34) \\

Object-12 &$\timeform{21h10m27s.8}$& $\timeform{48D14'30".8}$ & 24.34${\pm}$0.10& 20.81${\pm}$0.03 & 20.39${\pm}$0.06 & 19.35${\pm}$0.06  \\
~&~&~& (0.75) & (0.47) & (0.50) & (0.20) \\

Object-13 &$\timeform{21h11m23s.9}$& $\timeform{48D11'43".7}$ & 22.10${\pm}$0.05& 19.34${\pm}$0.03 & 19.28${\pm}$0.06 & 18.53${\pm}$0.06  \\
~&~&~& (0.06) & (0.66) & (0.09) & (0.69) \\

Object-14 &$\timeform{21h11m37s.4}$& $\timeform{48D21'45".5}$ & 22.95${\pm}$0.05& 19.93${\pm}$0.03 & 19.40${\pm}$0.05 & 18.64${\pm}$0.03  \\
~&~&~& (0.75) & (0.47) & (0.50) & (0.20) \\

Object-15 &$\timeform{21h10m50s.9}$& $\timeform{48D32'18.5}$ & 21.39${\pm}$0.03& 18.54${\pm}$0.03 & 18.19${\pm}$0.05 & 17.55${\pm}$0.03  \\
~&~&~& (0.34) & (0.02) & (0.93) & (0.04) \\

Object-16 &$\timeform{21h10m21s.1}$& $\timeform{48D22'20".1}$ & 21.48${\pm}$0.04& 19.04${\pm}$0.03 & 18.73${\pm}$0.05 & 18.21${\pm}$0.04  \\
~&~&~& (0.70) & (0.61) & (0.85) & (0.65) \\

\hline\hline

\end{longtable} 
%%%%%%%%%%%%%%%%%%%%%%%%%%%%%%%%%%%%%%%%%%%%%%%%%%%%%%%%%%%%%%%%%%%%%%%%%

%%%%%%%%%%%%%%%%%%%%%%%%%%%%%%%%%%%%%%%%%%%%%%%%%%%%%%%%%%%%%%%%%%%%%%%%%
\begin{longtable}{ccccc}
  \caption{Estimated parameters for W UMa-like objects assuming they are dwarfs.}\label{tab:LT5}
 \hline\hline\
Object & Spectral type & Period & Amplitude & Distance\\
~& ~&[days]& Primary{$\vert$}Secondary[\%] & [kpc]   \\

\hline
\endhead
Object-6& K5V/M2V& 0.309 & 2 & 0.7-1.7 \\
Object-7& K0V/M2V& 0.225 & 5$\vert$3 & 3.1 \\
Object-8& F8V/K2V or M2V/M5V& 0.254 & 3 & 1.6 or 0.4-1.1 \\
Object-9& G5V/K5V or M2V/M5V& 0.260 &12$\vert$7 & 6.7 or 1.3-3.3\\
Object-10& F5V/K2V or M2V/M5V& 0.346 &10 & 5.4 or 0.5-1.3 \\
Object-11& F8V/K2V or M2V/M5V& 0.358 &10 & 9.3 or 0.7-1.9 \\
Object-12& F0V/G8V & 0.362 &18$\vert$15& 5.1 \\
Object-13& F0V/K0V & 0.270 & 15$\vert$13 & 2.1 \\
Object-14& G5V/K5V or M0V/M5V& 0.259 &3 & 2.8 or 0.4-1.6 \\
Object-15& G0V/K7V or M0V/M5V& 0.368 &8$\vert$6 & 5.8 or 0.3-1.2 \\
Object-16& G5V/K5V or M0V/M5V& 0.252 &2& 2.5 or 0.5-1.9 \\

\hline\hline

\end{longtable} 
%%%%%%%%%%%%%%%%%%%%%%%%%%%%%%%%%%%%%%%%%%%%%%%%%%%%%%%%%%%%%%%%%%%%%%%%%

Object-6: The amplitude is approximately 2{\%}. The period is 0.309
days. Multicolor photometry suggests a K5/M2 star. If Object-6 is a
dwarf, the distance is 0.7-1.7 kpc.
 
Object-7: The primary minimum and the secondary minimum are
approximately 5{\%} and 3{\%}, respectively. The period is 0.225
days. Multicolor photometry suggests a K0/ M2 star. The distance of 3.1
kpc obtained from equation (A1) and equation (A2).

Object-8: The amplitude is approximately 3\%. The period is 0.254
days. Multicolor photometry suggests a F8/K2 or M2/M5 star. If Object-8
is a F8/K2 dwarf, the distance is 1.6 kpc. In the case of an M2/M5
dwarf, the estimated distance of 0.4-1.1 kpc is obtained from its expected absolute magnitude.

Object-9: The primary and secondary minimums are approximately 12\% and
7\%, respectively. The period is 0.260 days. Multicolor photometry
suggests a G5/K5 or an M2/M5 star. If Object-9 is a G5/K5 dwarf, the
distance is 6.7 kpc. In the case of an M2/M5 dwarf, the estimated
distance is 1.3-3.3 kpc.

Object-10: The amplitude is approximately 10\%. The period is 0.346
days. Multicolor photometry suggests a F5/K2 or an M2/M5 star. If
Object-10 is a F5/K2 dwarf, the distance is 5.4 kpc. In the case of an
M2/M5 dwarf, the estimated distance is 0.5-1.3 kpc.

Object-11: The amplitude is approximately 10\%. The period is 0.358
days. Multicolor photometry suggests a F8/K2 or M2/M5 star. If Object-11
is a F8/K2 dwarf, the distance is 9.3 kpc. In the case of an M2/M5
dwarf, the estimated distance is 0.7-1.9 kpc.

Object-12: The primary and secondary minimums are approximately 18\% and
15\%, respectively. The period is 0.362 days. Multicolor photometry
suggests a F0/G8 star. If Object-12 is a dwarf, the distance is 5.1 kpc.

Object-13: The primary and secondary minimums are approximately 15\% and
13\%, respectively. The period is 0.270 days. Multicolor photometry
suggests a F0/K0 star. If Object-13 is a dwarf, the distance is 2.1 kpc.

Object-14: The amplitude is approximately 3\%. The period is 0.259
days. Multicolor photometry suggests a G5/K5 or an M0/M5 star. If
Object-14 is a G5/K5 dwarf, the distance is 2.8 kpc. In the case of an
M0/M5 dwarf, the estimated distance is 0.4-1.6 kpc.

Object-15: The primary and secondary minimums are approximately 8\% and
6\%, respectively. The period is 0.368 days. Multicolor photometry
suggests a G0/M5 star. If Object-15 is a G0/K7 dwarf, the distance is
5.8 kpc. In the case of an M0/M5 dwarf, the estimated distance is
0.3-1.2 kpc.

Object-16: The amplitude is approximately 2\%. The period is 0.252
days. Multicolor photometry suggests a G5/K5 star or an M0/M5 star. If
Object-16 is a G5/K5 dwarf, the distance is 2.5 kpc. In the case of an
M0/M5 dwarf, the estimated distance is 0.5-1.9 kpc.

%%%%%%%%%%%%%%%%%%%%%%%%%%%%%%%%%%%%%%%%%%%%%%%%%%%%%%%%%%%%%%%%%%%%%%%%%
\begin{figure}
  \begin{center}
\begin{tabular}{ccc}
\FigureFile(52mm,52mm){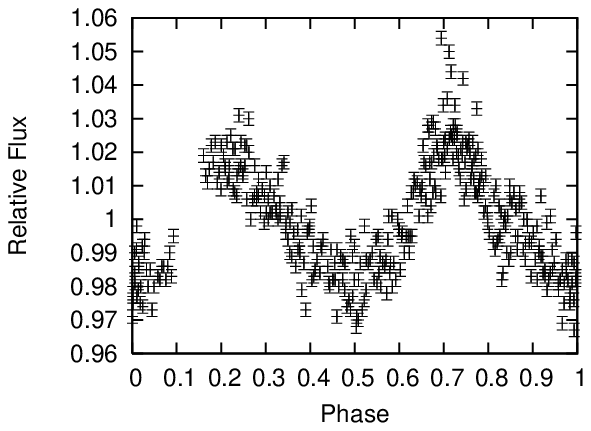}&\FigureFile(52mm,52mm){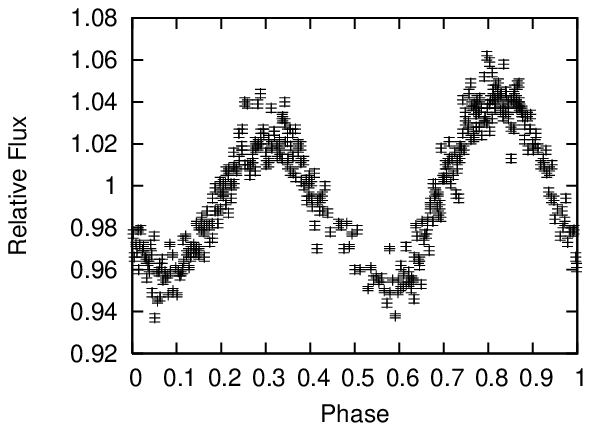}&\FigureFile(52mm,52mm){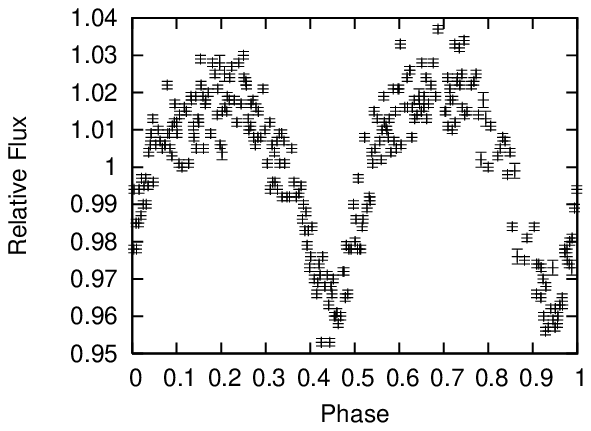}\\
~~~~~Object-6& ~~~~~Object-7&~~~~~Object-8\\
\FigureFile(52mm,52mm){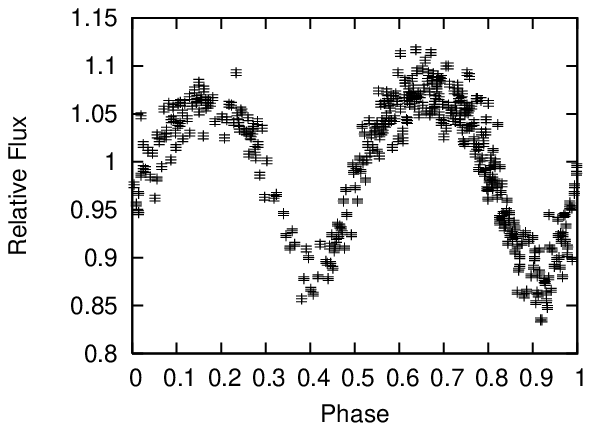}&\FigureFile(52mm,52mm){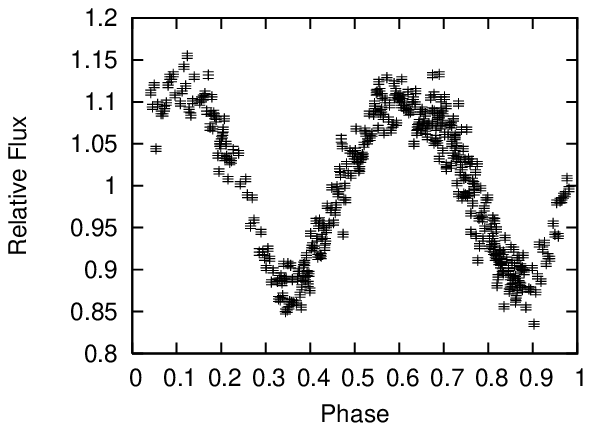}&\FigureFile(52mm,52mm){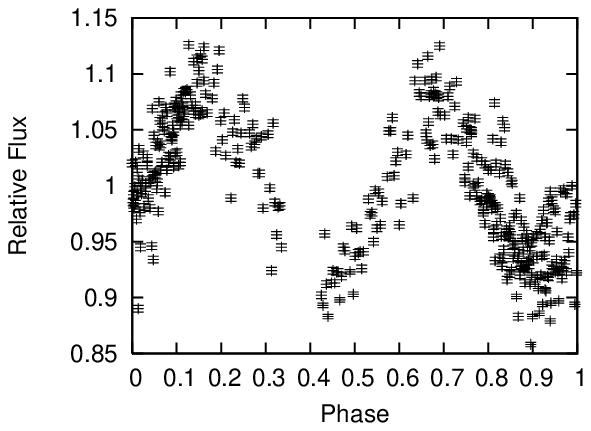}\\
~~~~~Object-9& ~~~~~Object-10&~~~~~Object-11\\
\FigureFile(52mm,52mm){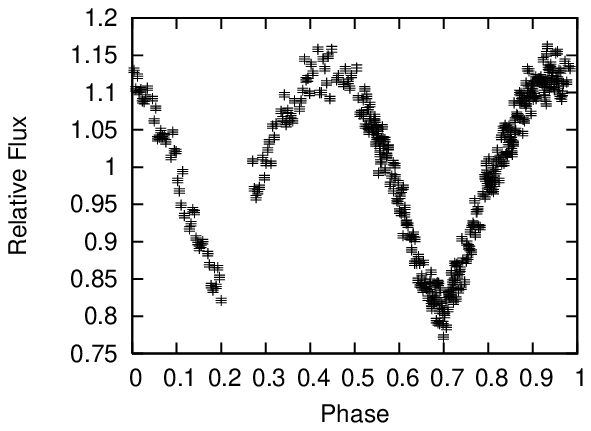}&\FigureFile(52mm,52mm){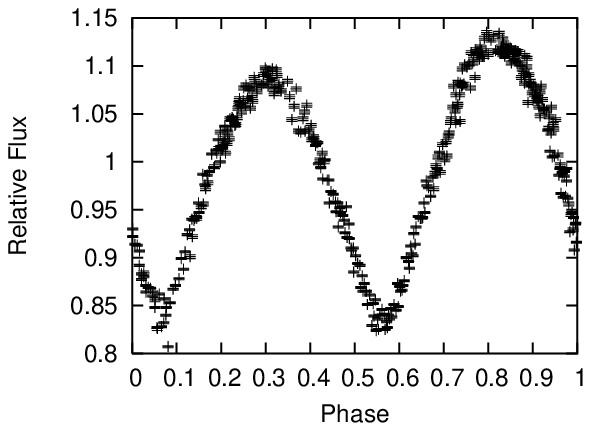}&\FigureFile(52mm,52mm){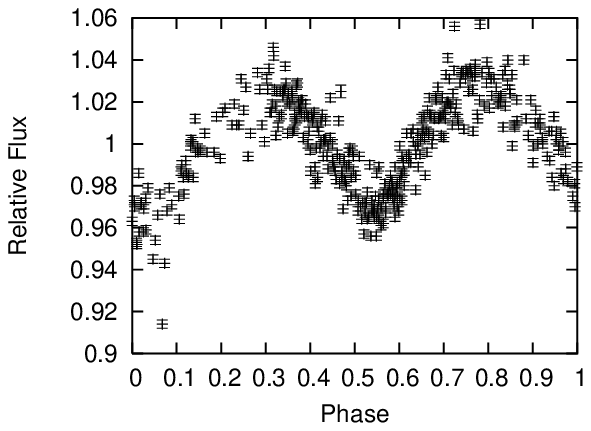}\\
~~~~~Object-12& ~~~~~Object-13&~~~~~Object-14\\
\FigureFile(52mm,52mm){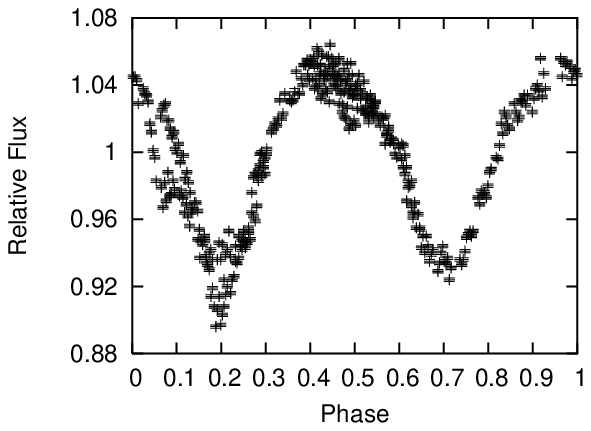}&\FigureFile(52mm,52mm){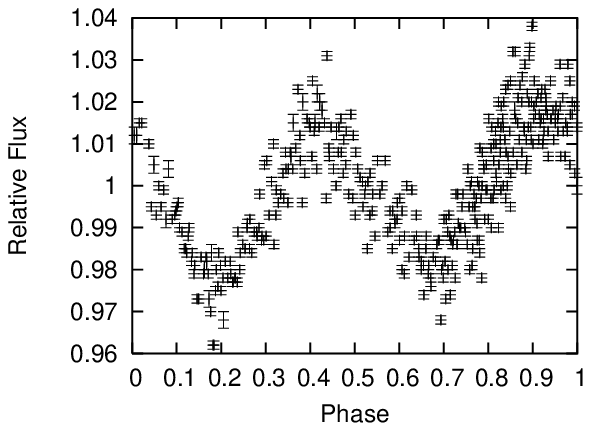}&\~~\\
 ~~~~~Object-15& ~~~~~Object-16&~~~~~\\

\end{tabular}
  \end{center}
  \caption{ Light curves of W UMa objects.}\label{fig:label7}
\end{figure}
%%%%%%%%%%%%%%%%%%%%%%%%%%%%%%%%%%%%%%%%%%%%%%%%%%%%%%%%%%%%%%%%%%%%%%%%%

%%%%%%%%%%%%%%%%%%%%%%%%%%%%%%%%%%%%%%%%%%%%%%%%%%%%%%%%%%%%%%%%%%%%%%%%%

%%%%%%%%%%%%%%%%%%%%%%%%%%%%%%%%%%%%%%%%%%%%%%%%%%%%%%%%%%%%%%%%%%%%%%%%%

\end{document}